\documentclass[11pt,preprintnumbers,aps,amssymb,nofootinbib,amsmath,superscriptaddress,notitlepage]
{revtex4-1}
\usepackage{epsfig,epsf}
\usepackage{comment}
\usepackage{bm} 
\usepackage{color} 
\usepackage{slashed} 
\usepackage{relsize}	
\usepackage{soul} 
\usepackage{hyperref}
\usepackage{tensor} 
\newcommand{\beq}{\begin{equation}}
\newcommand{\beql}[1]{\begin{equation}\label{#1}}
\newcommand{\eeq}{\end{equation}}
\def\bal#1\gal{\begin{align}#1\end{align}}
\newcommand{\ball}[1]{\bal\label{#1}}
\newcommand{\eq}[1]{(\ref{#1})}
\newcommand{\fig}[1]{Fig.~\ref{#1}}
\renewcommand{\sec}[1]{Sec.~\ref{#1}}
\DeclareMathOperator{\sgn}{sgn}

\DeclareMathOperator{\im}{\mathrm{Im}}
\DeclareMathOperator{\real}{\mathrm{Re}}

\renewcommand{\b}[1]{{\bm #1}} 
\newcommand{\unit}[1]{\hat {{\bm #1}}} 

\newcommand{\e}{\varepsilon}

\newcommand{\jpsi}{J\mskip -2mu/\mskip -0.5mu\psi} 

\begin{document}

\title{Magneto-rotational dissociation of heavy hadrons\\ in Relativistic Heavy-Ion Collisions}

\author{Kirill Tuchin}

\affiliation{
Department of Physics and Astronomy, Iowa State University, Ames, Iowa, 50011, USA}

\date{\today}

\begin{abstract}

A heavy hadron traversing Quark-Gluon Plasma and dragged along in rotational motion is subject to the Lorentz and centrifugal forces. The Lorentz force, sourced by the valence quarks of heavy-ions, possesses  the electric and magnetic components in the hadron comoving frame. The electric component renders the hadron unstable by empowering one of its quarks to tunnel through the potential barrier. Assuming that the magnetic field is parallel to the plasma vorticity, the hadron dissociation probability is computed using the Imaginary Time Method and  
is found to strongly depend on the sign of the quark electric charge. The dissociation probability monotonically increases as a function of vorticity for negative electric charges, whereas for positive charges it exhibits a minimum at a finite value of  vorticity. The dissociation probability for the negative charges is larger than for the positive ones at the same magnetic field and vorticity.  In relativistic heavy-ion collisions this implies lower abundance of the negatively charged hadrons as compared to the positively charges ones. This effect is significant at moderate collision energies where the plasma vorticity is comparable or larger than the synchrotron frequency. 

\end{abstract}

\maketitle

\section{Introduction}\label{sec:a}

A heavy quarkonium, or any other hadronic state, immersed into Quark-Gluon Plasma (QGP) dissociates as a result of the Debye screening \cite{Matsui:1986dk,Mocsy:2013syh} and induction of the imaginary part of the potential \cite{Digal:2001ue,Wong:2004zr,Mocsy:2005qw,Alberico:2006vw,Cabrera:2006wh}. If QGP were contained in an externally controlled environment this would have been its most clear signature. In reality, the correlation between the bound state dissociation and the hadronic spectra in relativistic heavy-ion collisions is ambiguous as there are other factors that contribute to the dissociation  \cite{Brambilla:2010cs}. In particular, it was proposed in \cite{Marasinghe:2011bt} that dissociation of hadronic states is also caused by the magnetic field sourced by the valence electric charges of the colliding ions \cite{Kharzeev:2007jp,Skokov:2009qp,Voronyuk:2011jd,Ou:2011fm,Deng:2012pc,Tuchin:2013apa}. This is most clearly seen in the hadron comoving frame where a quark can tunnel through the potential barrier of the binding force under the influence of the electric field. Considering that the electric field in the plasma comoving frame vanishes, the electric field in the hadron comoving frame equals the cross-product of the velocity of the hadron with respect to plasma and the magnetic field. The dissociation  probability  increases with the electric field and decreases with the magnetic one \cite{Marasinghe:2011bt}. The dissociation is just one of many effects that magnetic field has on hadrons. The quantum mechanical problem of the heavy hadron  in the magnetic field, in the context of relativistic heavy-ion collisions,  was examined in a number of publications \cite{Marasinghe:2011bt,Tuchin:2011cg,Machado:2013rta,Alford:2013jva,Cho:2014loa,Dudal:2014jfa,Sadofyev:2015hxa,Yoshida:2016xgm,Singh:2017nfa,Dutta:2017pya,Suzuki:2016fof,Liu:2018zag,Iwasaki:2018pby}.

It has recently been observed that some of the orbital angular momentum of the relativistic heavy-ions is transferred to the QGP conferring on it finite space-time dependent vorticity, which points in the same direction as the magnetic field \cite{Csernai:2013bqa,Csernai:2014ywa,Becattini:2015ska,Deng:2016gyh,Jiang:2016woz,Kolomeitsev:2018svb,Deng:2020ygd,Xia:2018tes}. The combined effect of rotation and magnetic field on QGP has been recently discussed in  \cite{Chen:2015hfc,Mameda:2015ria}. The goal of this paper is to investigate the effect of rotation on hadron dissociation. The analysis is most conveniently performed in the hadron comoving frame rotating with respect to the laboratory frame. 
As in \cite{Marasinghe:2011bt} our analysis will ignore the complexity of the electromagnetic field and instead focus on a qualitative analysis that assumes that in the plasma comoving frame there is only uniform time-independent magnetic field $\b B$, while the electric field vanishes as dictated by the Ohm's law. It is further assumed that as the heavy hadron traverses the plasma with the relative translation velocity $\b V$ it is dragged along by the plasma to rotate with uniform, time-independent and parallel to the magnetic field angular velocity/vorticity $\b\Omega$. In the hadron comoving frame then, the bound state is subject to the mutually orthogonal and constant electric and magnetic fields, the former is given by  $\b E=\b V\times \b B$ while the later is invariant under the reference frame transformations in the non-relativistic approximation that we adopt throughout the paper.

The dissociation of a bound state occurs due to the tunneling of one of its quarks under the action of the electric field. Its probability can be computed in the quasi-classical approximation using the Imaginary Time Method \cite{popov-review}. The idea is to consider the quasi-classical trajectories in the classically forbidden region under the potential barrier as classical motion in the Euclidean time $\tau =it$. The dissociation probability $w$ is related to the imaginary part of the restricted action $W$ (defined below) along the extremal trajectory in the following way:
\ball{a2}
w= \exp(-2\im W)\,.
\gal
It can be computed analytically under a few assumptions. Firstly, it is assumed that a hadron is made up of a heavy and a light quarks (e.g.\ $D$-meson)  and that motion of the heavy-quark with respect to the hadron center-of-mass is negligible. This circumvents a major difficulty in solving the two-body problem with its inseparable   center-of-mass motion in the magnetic field in an inertial frame, let alone in the rotating frame. Secondly, the binding potential is assumed to be short-range, i.e.\ the hadron radius is much smaller than the radius of the lowest Landau orbit. The finite radius correction gives a slow varying logarithmic contribution to $W$. It is important for quantitative analysis, but can be neglected in a qualitative discussion \cite{Marasinghe:2011bt}.  Thirdly, we assume that motion of the quark along the quasi-classical trajectory is non-relativistic, which is an accurate approximation for  most phenomenological applications in heavy-ion collisions \cite{Marasinghe:2011bt}. 

The paper is structured as follows. The equations of motion are derived in \sec{sec:b} and solved in \sec{sec:c}. The Imaginary Time Method is then employed to identify the extremal sub-barrier trajectory. The corresponding restricted action is computed in \sec{sec:d}. The main result is Eqs.~\eq{d5} and \eq{c34} and \fig{fig:F}. In \sec{sec:d}  the peculiar features of the dissociation probability as a function of $B$ and $\Omega$  are discussed. The two main observations are: (i) the dissociation probability $w$ of negative electric charges increases as a function of $\Omega$ and (ii) $w$ of positive electric charges has minimum at a  certain angular velocity not exceeding the synchrotron frequency. The dissociation is thus sensitive to the sign of the light quark electric charge. These features can be seen in \fig{fig:W}. The increase of the dissociation probability with vorticity is the consequence of the centrifugal force, while the dependence on the electric charge sign stems from the different direction of rotation of positive and negative charges in the magnetic field as shown in \fig{fig:G}.  Phenomenological implications to heavy-ion collisions are examined in \sec{sec:s}.

\section{Equations of motion}\label{sec:b}

Equation of motion of a quark  of mass $m$ and electric charge $e$ in the electromagnetic field reads: 
\ball{c2}
\frac{du^\mu}{ds}+\Gamma\indices{^\mu _\nu _\lambda}u^\nu u^\lambda= \frac{e}{mc^2}F^{\mu\nu}u_\nu\,,
\gal
where $u^\mu = dx^\mu/ds$ is four-velocity. In a frame  rotating with angular velocity $\b \Omega= \Omega \unit z$ with respect to the laboratory frame (which is the center-of-mass frame of the heavy-ion collision)  the invariant interval is
\ball{b8}
ds^2= g_{\mu\nu}x^\mu x^\nu= [c^2-\Omega^2(x^2+y^2)]dt^2 -dx^2-dy^2-dz^2+2\Omega y dx dt-2\Omega x dy dt\,.
\gal
The corresponding non-vanishing Christoffel symbols are
\ball{c4}
\Gamma\indices{^x _t _t}=-x\Omega^2/c^2\,,\qquad \Gamma\indices{^x _t _y}= \Gamma\indices{^x _y_t}=-\Omega/c\,,\qquad 
\Gamma\indices{^y _t _t}=-y\Omega^2/c^2\,,\qquad \Gamma\indices{^y _t _x}= \Gamma\indices{^y _x_t}=\Omega/c\,.
\gal

Let $\b B= B\unit z$ and $\b E= E\unit y$ be the magnetic and electric fields in the hadron comoving frame. We assume that $\Omega, E, B$ are positive. However, the charge $e$ can be either positive or negative. The electric and magnetic fields are related to the components of the field strength tensor $F_{\mu\nu}$ as 
\ball{b10}
E_i= F_{0i}\,, \quad B^i=-\frac{1}{2\sqrt{\det \gamma}}\e^{ijk}B_{jk}\,,\quad  B_{ij}= F_{ij}\,,
\gal
where $\gamma_{ij}=-g_{ij}+g_{0i}g_{0j}/g_{00}$ is the three-dimensional spatial metric. In the non-relativistic approximation $\gamma_{ij}\approx \delta_{ij}$ up to the terms of order $1/c^2$. Thus, the non-vanishing components of the field strength tensor are $F_{02}=-F_{20}=E$ and $F_{21}=-F_{12}=B$. The indices are raised using the contravariant tensor $g^{\mu\nu}= g_{\mu\nu}+\mathcal{O}(1/c^2)$ with the result 
\ball{b13}
F^{\mu\nu}= \left(\begin{array}{cccc}0 & x\Omega B/c & -E+y\Omega B /c& 0 \\-x\Omega B/c & 0 & -B -y \Omega E/c & 0 \\E-y\Omega B/c & B+y\Omega E/c & 0 & 0 \\0 & 0 & 0 & 0\end{array}\right)+\mathcal{O}(1/c^2)\,.
\gal

The covariant components of the four-velocity  are 
\ball{b15}
u_t\approx u^t\approx 1\,,\quad  u_x\approx (-v^x+y\Omega)/c\,,\quad  u_y\approx (-v^y-x\Omega)/c\,,\quad u_z= -v^z/c\,.
\gal
Substituting \eq{b13},\eq{b15} and $s\approx c t$ into \eq{c2} we obtain the equations of motion in the hadron comoving frame at the leading non-relativistic order 
\bal
&\dot v^x-2\Omega v^y-x\Omega^2= \frac{eB}{mc}v^y \,, \label{c7}\\
&\dot v^y+2\Omega v^x-y\Omega^2=\frac{eE}{m}-\frac{eB}{mc}v^x\,,\label{c8}\\
&\dot v^z=0\,.\label{c9}
\gal
where the dotted symbols are time-derivatives.

Eq.~\eq{c2} can be obtained as the Euler-Lagrange equation  from  the Lagrangian 
\ball{b1}
L=-\left(mc+\frac{e}{c}A_\mu u^\mu\right) \frac{ds}{dt}\,.
\gal
Choosing the symmetric gauge
\ball{b3}
A_\mu=\left(-Ey,\frac{1}{2}By,-\frac{1}{2}Bx,0\right)\,,
\gal
substituting \eq{b8} and expanding in inverse powers of $c$  one finds the non-relativistic expression for the Lagrangian 
\ball{b4}
L= -mc^2 + \frac{m\b v^2}{2} +m\Omega(xv^y-y v^x)  + \frac{m}{2}(x^2+y^2)\Omega^2
+\frac{e}{2c}B(x v^y-yv^x)+eEy+\mathcal{O}(1/c^2)\,,
\gal 
which yields the equations of motion \eq{c7}-\eq{c9}.  From now on I will use the natural units with $c=\hbar=1$. 

One may question the expediency of using the relativistic approach in this section, since  \eq{c7}-\eq{c9}  can be easily derived from the non-relativistic equations of motion in the rotating frame. The advantage of the covariant approach is that it can be easily generalized to more complicated field configurations and allows systematic computation of the relativistic corrections, and eventually solving the fully relativistic problem (perhaps numerically).

\section{Sub-barrier trajectory}\label{sec:c}

The classical sub-barrier  trajectory that minimizes the restricted action  is a solution to Eqs.~\eq{c7}-\eq{c9}. It  starts at the imaginary time $t_0$, ends at $t=0$ and satisfies the boundary conditions \cite{popov-review}:
\bal
\b r(t_0)&= 0\,,\label{c24}\\
\im \b r(0)&= \im \b v(0)=0\,, \label{c25}\\
\frac{1}{2}m v^2(t_0)&=\e_0-mc^2 = -\e_b\,. \label{c26}
\gal
where $\e_b>0$ is the hadron binding energy. Eq.~\eq{c26} determines the initial time $t_0$ of the sub-barrier motion.

Multiplying \eq{c8} by $i$ and adding \eq{c7} we obtain an equation for $\xi = x+iy$: 
\ball{c16}
\ddot \xi + i(2\Omega+\omega_B)\dot \xi -\Omega^2\xi = i\omega_E \,,
\gal
where we denoted 
\ball{c11}
\omega_E= \frac{eE}{m}\,,\qquad \omega_B= \frac{eB}{m}\,.
\gal
The general solution to \eq{c16} is
\ball{c18}
\xi = C_1 e^{-it\omega_+}+C_2e^{-i t\omega_-}- \frac{i\omega_E}{\Omega^2}\,,
\gal
where the characteristics frequencies are 
\ball{c19}
\omega_\pm = \Omega+\frac{\omega_B}{2}\pm \sqrt{\Omega \omega_B+\frac{\omega_B^2}{4}}\,,
\gal
and $C_1$ and $C_2$ are complex constants. We are going to derive the extremal trajectory for real $\omega_\pm$ and then analytically continue it to complex values (at $-4\Omega < \omega_B<0$). Accordingly, taking the real and imaginary parts of \eq{c18} and introducing real constants $A_{1,2}$, $B_{1,2}$ we obtain
\bal
x&= A_1\cos(\omega_+ t)-A_2\sin(\omega_+ t)+B_1\cos(\omega_- t) -B_2\sin( \omega_- t) \,,\label{c20}\\
y&=-A_1\sin(\omega_+t)-A_2\cos(\omega_+ t)-B_1\sin (\omega_- t)-B_2\cos(\omega_- t) 
-\frac{\omega_E}{\Omega^2}\,.\label{c21}
\gal

In the classically forbidden region, it is expedient to parameterize the trajectory  in terms of the real Euclidean time $\tau =it$. Applying the initial condition \eq{c24} one derives 
\bal
x(\tau)&=\frac{i\omega_E}{\Omega^2\sinh[(\omega_+-\omega_-)\tau_0]}\left\{ \sinh(\omega_-\tau_0)\sinh(\omega_+\tau)
-\sinh(\omega_+\tau_0)\sinh(\omega_-\tau)\right\}
 \,, \label{c30}\\
y(\tau)&=\frac{\omega_E}{\Omega^2}\left\{
-\frac{\sinh(\omega_-\tau_0)\cosh(\omega_+\tau)}{\sinh[(\omega_+-\omega_-)\tau_0]}
+\frac{\sinh(\omega_+\tau_0)\cosh(\omega_-\tau)}{\sinh[(\omega_+-\omega_-)\tau_0]}-1
 \right\}\,,\label{c31}
\gal
where $\tau_0\le \tau \le 0$ and $\tau_0<0$. 
  Continuing \eq{c30} and \eq{c31}  to the complex values of the characteristic frequencies $\omega_\pm$ furnishes the desired extremal trajectory for any $\omega_E$, $\omega_B$,  and $\Omega$.  The extremal sub-barrier trajectory lies entirely in the $xy$-plane because the boundary condition \eq{c25} applied to the solution of \eq{c9}: $z= v_z(t-t_0)=-iv_{z0}(\tau-\tau_0)$  implies $v_{z0}=0$.

\section{Dissociation probability}\label{sec:d}

\begin{figure}[ht]
      \includegraphics[height=6cm]{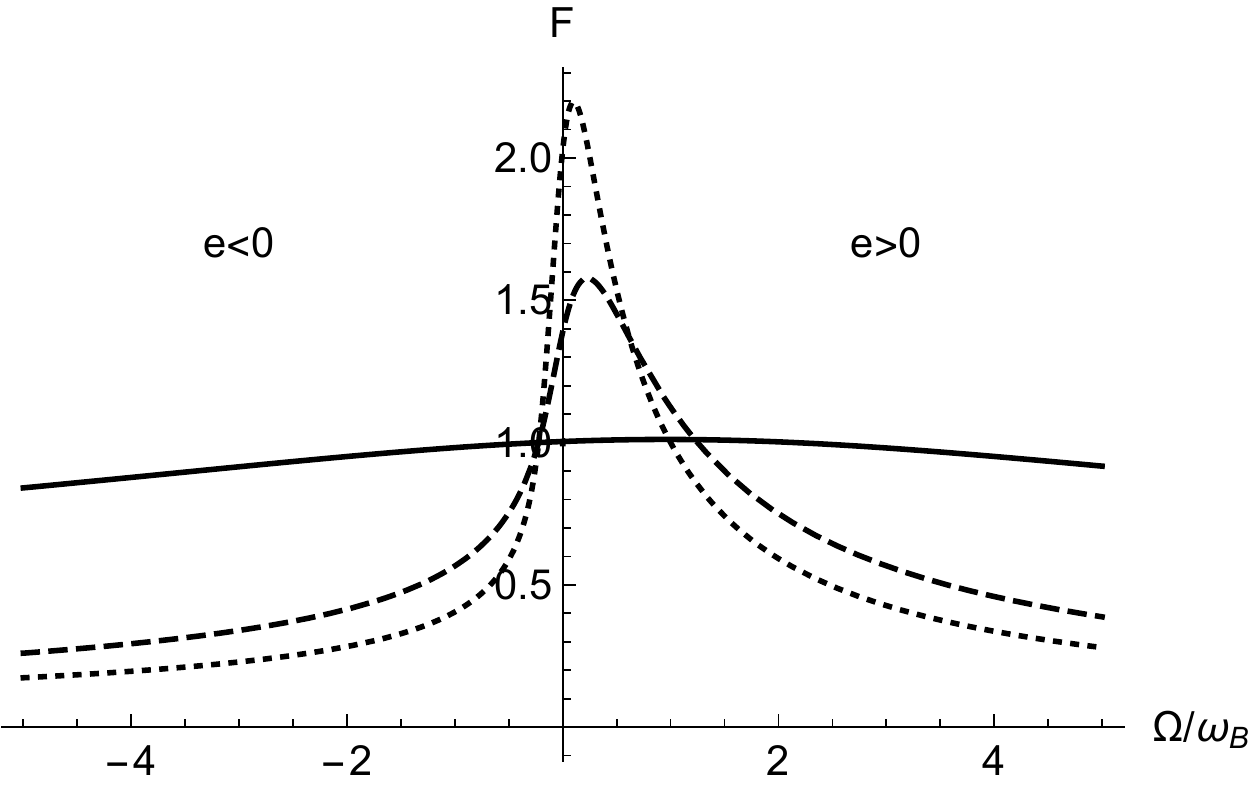} 
  \caption{$F=\im W(3\omega_B^3/m \omega_E^2\gamma^3)$ versus $\Omega/\omega_B$ for $\gamma=1/3$ (solid line), $\gamma=3$ (dashed line) and $\gamma=5$ (dotted line). Notice that at given $\Omega$,  $F$ is larger for $e>0$ ($\omega_B>0)$ than for $e<0$ ($\omega_B<0$). }
\label{fig:F}
\end{figure}

The restricted action $W$ appearing in \eq{a2} is given by
\ball{d2}
W= \int_{t_0}^0(L+\e_0)dt- \b p\cdot \b r|_{t=0}\,.
\gal
 Substituting the Lagrangian \eq{b4} and the trajectory \eq{c30},\eq{c31} one obtains 
\ball{d5}
W=i\frac{m\omega_E^2}{8\Omega^4}\frac{\omega_+-\omega_-}{\sinh^2[(\omega_+-\omega_-)\tau_0]}
\big\{-2\tau_0(\omega_+-\omega_-)+2\tau_0\omega_+\cosh(2\omega_-\tau_0)-2\omega_-\tau_0\cosh(2\omega_+\tau_0)\nonumber
 \\
-\sinh[2(\omega_+-\omega_-)\tau_0]-\sinh(2\omega_-\tau_0)+\sinh(2\omega_+\tau_0)\big\}\,.
\gal
Eq.~\eq{c26} determines implicitly the initial instant of the sub-barrier motion  $\tau_0$: 
\ball{c34}
\gamma^2= \frac{\omega_B^2}{\Omega^4\sinh^2[(\omega_+-\omega_-)\tau_0]}
\big\{
2\omega_+\omega_-\sinh(\omega_-\tau_0)\sinh(\omega_+\tau_0)\cosh[(\omega_+-\omega_-)\tau_0]
&\nonumber\\
-\omega_+^2\sinh^2(\omega_-\tau_0)-\omega_-^2\sinh^2(\omega_+\tau_0)\big\}
\,,&
\gal
where we introduced a positive dimensionless  parameter \cite{Popov:1997-A,Popov:1998-A}
\ball{c35}
 \gamma= \sqrt{\frac{2\e_b}{m}}\frac{B}{E}\,.
\gal 
Eliminating $\tau_0$ from \eq{d5} and \eq{c34} and substituting into in \eq{a2} yields the desired dissociation probability. \fig{fig:F} displays the function $F$ defined as 
\ball{c40}
F=\frac{3|\omega_B|^3}{m \omega_E^2\gamma^3}\im W=\frac{3|e|E}{m^2}\left( \frac{m}{2\e_b}\right)^{3/2}\im W\,,
\gal 
which is the imaginary part of $W$ normalized to unity in the limit $\omega_B\to 0$ and $\Omega\to 0$, see \eq{d33}. We observe that $F$  falls off with $\Omega$ indicating that the dissociation probability $w$ increases with the angular velocity. This occurs because the centrifugal force makes the bound state less stable. 

\begin{figure}[ht]
      \includegraphics[height=4cm]{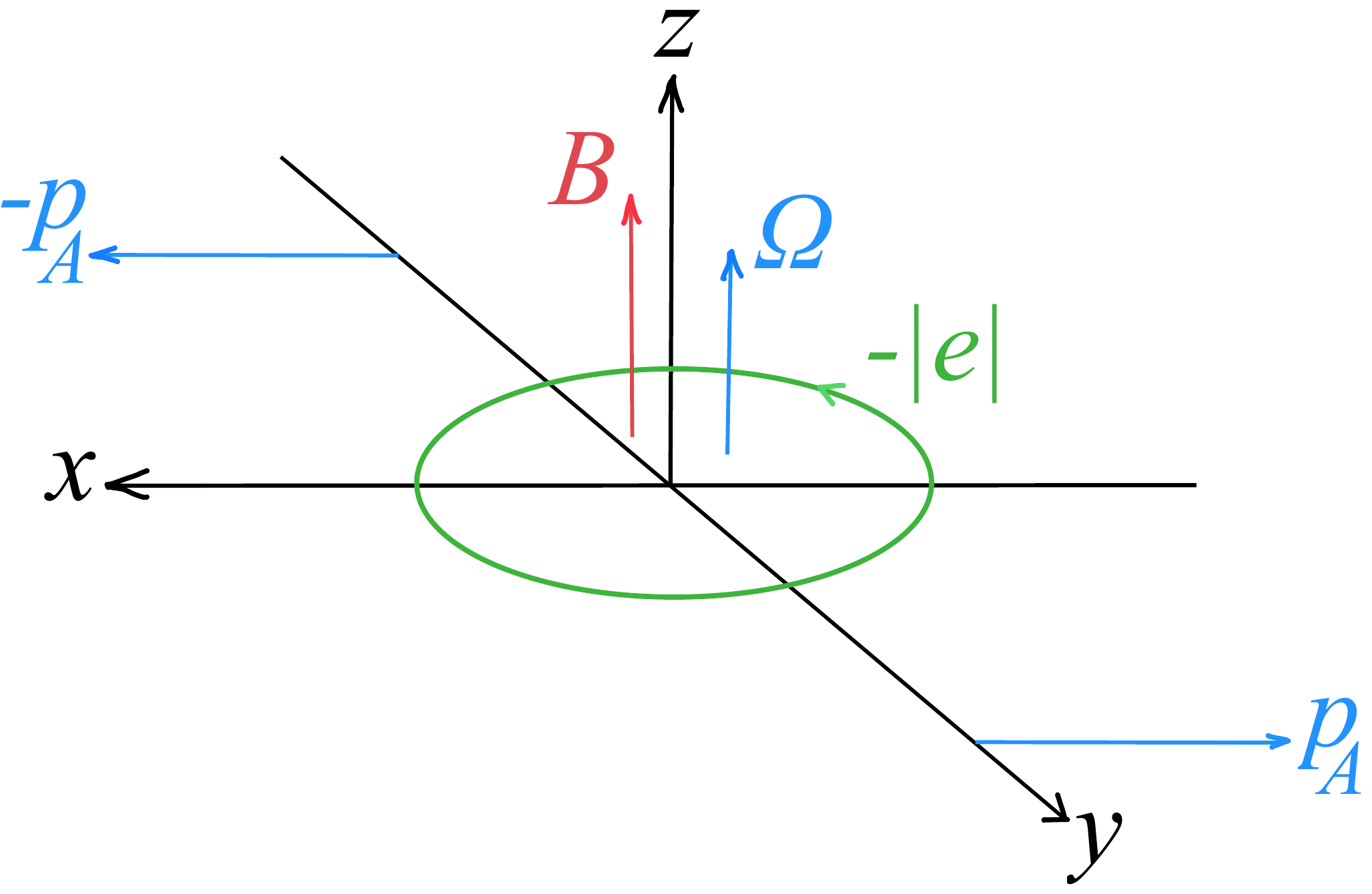} 
  \caption{A view from the laboratory frame: $\b p_A$ is the initial heavy-ion momentum, $\b B$ is the magnetic field of the valence charges, $\b \Omega$ is the plasma vorticity, the circle is the classical trajectory of a negatively charged quark at vanishing vorticity, a positive charge would move in the opposite direction. }
\label{fig:G}
\end{figure}

One can also see that at a given $\Omega$ positive charges have larger $F$ hence smaller dissociation probability $w$ than the negative charges. In particular, the maximum of $F$, corresponding to the minimum of $w$, occurs for positive charges at a certain angular velocity $\Omega < \omega_B$. To understand this effect, observe that in the laboratory frame a negative charge rotates counterclockwise  in the magnetic field, whereas a positive charge rotates in the opposite direction, see \fig{fig:G}. Therefore, in the hadron comoving frame, rotating with angular velocity $\Omega$, the effective synchrotron frequency of the negative charge $-|e|$ appears to be smaller (in absolute value) than the effective synchrotron frequency of a positive charge $|e|$. This implies that  the negative charge is perceived in this frame to be a subject to a weaker magnetic field than the positive charge. Since the dissociation probability decreases with the magnetic field, it is larger for negative charges than for positive ones.

Consider now several limiting cases that illuminate the role of various parameters in the dissociation dynamics. It is useful to define the dimensionless parameters 
\ball{c50}
\eta= -\omega_B\tau_0\,,\quad \kappa=\Omega/\omega_B\,.
\gal 
Since $\tau_0<0$, $\eta>0$ for positive charges, and $\eta<0$ negative ones.  In either case $\im W$ is positive. 

\subsection{No rotation $\Omega=0$.}

In the limit $\Omega=0$  Eqs.~\eq{d5}  and \eq{c34}  reduce to the known results \cite{Popov:1997-A,Popov:1998-A}  
\bal
2\im W&=\frac{m^2E^2}{|e|B^3}\eta^2\left(\coth\eta-\eta\coth^2\eta+\eta\right)\,,\label{d11}\\
\gamma^2&= \eta^2-\left(\eta\coth\eta-1\right)^2\,\label{d10}
\gal
that give the dissociation probability in mutually perpendicular electric and magnetic fields in a non-rotating frame.

\subsection{Strong electric field $E\gg B\sqrt{2\e_b/m}$.}

In such strong electric fields $\gamma\ll 1$ which is achieved if $|\eta|\ll1 $. Keeping $\kappa$ fixed and  expanding  \eq{c34} and \eq{d5}  we derive 
\bal
\sgn(e)\eta&= \gamma+ \frac{1}{18}\left(1+\frac{4\Omega}{\omega_B}-\frac{2\Omega^2}{\omega_B^2}\right)\gamma^3+\mathcal{O}(\gamma^5)\,,\label{d32}\\
2\im W&\approx \frac{2m^2}{3|e|E}\left(\frac{2\e_b}{m}\right)^{3/2}
\left[ 1+ \frac{1}{30}\left(1+\frac{4\Omega}{\omega_B}-\frac{2\Omega^2}{\omega_B^2}\right)\gamma^2+\mathcal{O}(\gamma^4)\right]\,.
\label{d33}
\gal
Actually, at large angular velocity $\Omega\gg |\omega_B|$ these equations are valid only if a stronger condition $\gamma\Omega/|\omega_B|\ll 1$ is satisfied. At small angular velocity $\im W$ increases with $\gamma$ i.e.\ as $E$ decreases, whereas at larger $\Omega$ the effect is opposite: the larger $E$ corresponds to the larger $\im W$. Thus, as a function of  $\Omega$, the dissociating probability has minimum at $\Omega=|\omega_B|$, which decreases as $\gamma$ increases, as can be seen in \fig{fig:F}.

One may notice that \eq{d33} contains a term linear in $\Omega/\omega_B$. This term has opposite sign for positive and negative charges. The difference between the  negative and positive charges in the same magnetic field is
\ball{d35}
2\im W(e>0)-2\im W(e<0)=  \frac{8m^3\Omega B }{45e^2 E^3}\left(\frac{2\e_b}{m}\right)^{5/2}\,.
\gal
This indicates that the negative charge has higher tunneling probability than the positive one, see \fig{fig:W}. While this effect is small when $\gamma\ll 1$, it is enhanced at large $\gamma$ as we proceed to demonstrate.

\subsection{Weak electric field $E\ll B\sqrt{2\e_b/m}$.}

Weak electric field corresponds to $\gamma\gg 1$ which requires $|\eta|\gg 1$. Consider first $\kappa>-1/4$ at which $\omega_\pm$ are real. Eq.~\eq{c34} can then be approximated as 
\ball{d36}
\gamma^2=\frac{2}{\kappa^4}\left[(2\kappa+1)\sqrt{4\kappa+1}-4\kappa-1\right]\exp\left\{ |\eta|(2\kappa+1-\sqrt{4\kappa+1})\right\}\,.
\gal 
Substituting this into \eq{d5} and keeping only the leading terms we obtain
\ball{d38}
2\im W= \frac{m^2E^2}{|e|B^3}\frac{\gamma^2\sqrt{1+4\kappa}}{4\left[(2\kappa+1)\sqrt{4\kappa+1}-4\kappa-1\right]}\ln\left( \frac{\gamma^2\kappa^4}{2\left[(2\kappa+1)\sqrt{4\kappa+1}-4\kappa-1\right]}\right)\,.
\gal

At $\kappa<-1/4$ \eq{c34} can be  equivalently  (i.e.\ making no approximations) cast in the following form:
\ball{d40}
\gamma^2= \frac{1}{2\kappa^4\sin^2(\eta r)}\bigg\{ \real\, (2\kappa+1+ir)^2\sinh^2[\eta(\kappa+1/2-ir/2)]\nonumber\\
-4\kappa^2\left| \sinh[\eta(\kappa+1/2-ir/2)]\right|^2\cos(\eta r)
\bigg\}\,,
\gal
where we denoted $\sqrt{1+4\kappa}=ir$. The presence of periodic functions of $\eta$ in \eq{d40} indicates that $\gamma(\eta)$ is not a one-to-one function for all $\eta$'s.  Rather it has to be constrained to the physical interval $0\le \sgn(e)\eta<\pi/r$, which contains the correct limit \eq{d32}. Inverting \eq{d40} in the specified interval, one obtains $\eta(\gamma)$ which, upon substitution into \eq{d5}, yields $W$. The numerical results are shown in \fig{fig:F}. 

Taking now $|\eta|\gg 1$ one can obtain the desired approximation. However, since the result is bulky, consider further the limit $\kappa\ll -1/2$ which corresponds to large angular velocity (in the opposite direction with respect to the synchrotron frequency). The result  is
\ball{d42}
2\im W= \frac{m^2E^2}{|e|B^3}\frac{\gamma^2}{2|\kappa|}\ln\left (\gamma^2|\kappa|^{5/2}\right)\,.
\gal

\section{Phenomenological applications to heavy-ion collisions}\label{sec:s}

In heavy-ion collisions the electromagnetic field and vorticity have a complicated spatial and temporal dependence. However, analytical computation of the magneto-rotational dissociation is possible only in a few relatively simple models of the binding potential, electromagnetic field and vorticity. More realistic models have to be tackled numerically. Meanwhile, the results of the previous section can be used for a qualitative discussion of the general features of the vorticity dependence of the dissociation probability of heavy hadrons. This provides a benchmark for the future phenomenology of the magneto-rotational effect.  

\begin{figure}[ht]
      \includegraphics[height=5cm]{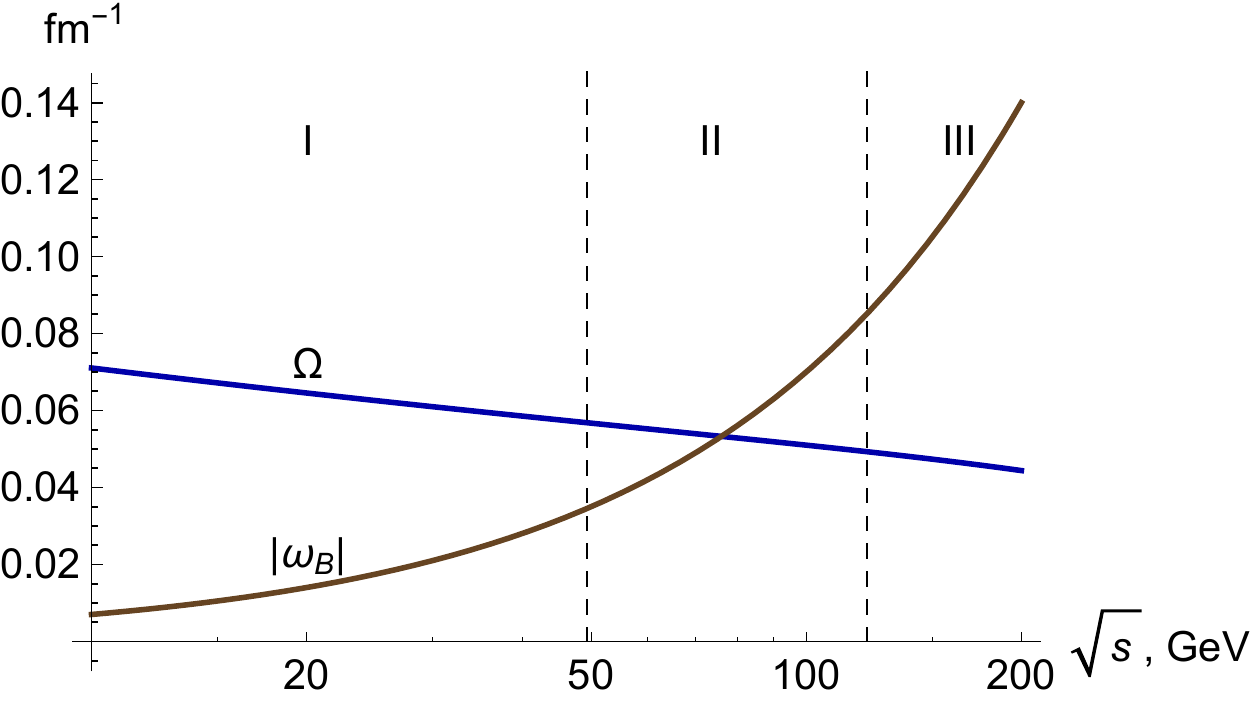} 
  \caption{The average vorticity of Quark-Gluon Plasma $\Omega$ and the typical synchrotron frequency $\omega_B$ versus the collision energy $\sqrt{s}$ (per a pair of nucleons) at $b=10$~fm immediately after the collision.  The vorticity is  gleaned from \cite{Huang:2020xyr}. $\omega_B$ is computed for $u$ quark with  mass $m=gT/\sqrt{3}$, where $g$ is the strong coupling and $T$ is temperature. }
\label{fig:omegas}
\end{figure}

First, we observe that the ratio $\kappa=\Omega/\omega_B$ strongly depends on the collision energy $\sqrt{s}$ and time, see \fig{fig:omegas}. Its magnitude decreases with the collision energy and  increases with time. At $\sqrt{s}\sim 50-130$~GeV $|\kappa|$ is of order unity, so that the magnetic field and  vorticity compete with each other \cite{Deng:2012pc,Deng:2016gyh,Deng:2020ygd} (region II in \fig{fig:omegas}). At higher energy the dissociation is dominated by the magnetic field $|\kappa|\ll 1$ (region III), while at lower energies by vorticity $|\kappa|\gg 1$ (region I). 

\begin{figure}[ht]
\begin{tabular}{cc}
      \includegraphics[height=5cm]{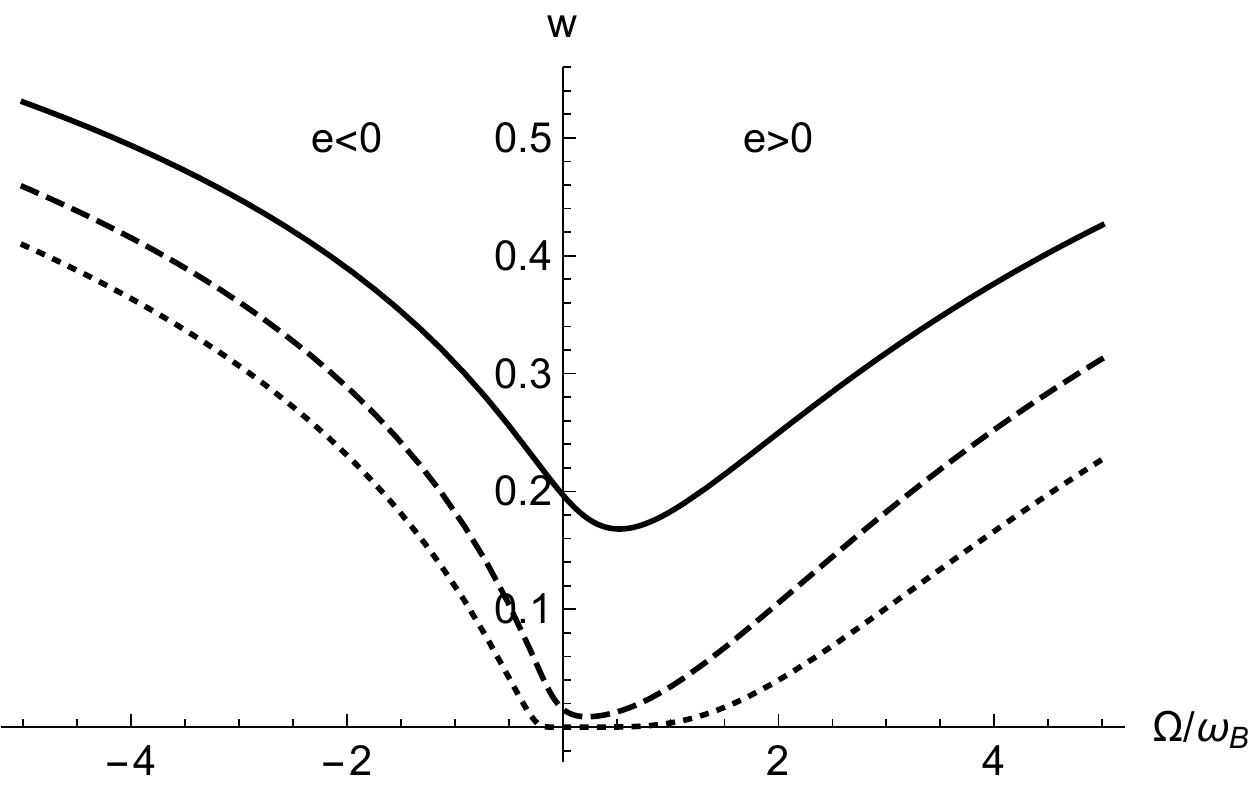} &
            \includegraphics[height=4.5cm]{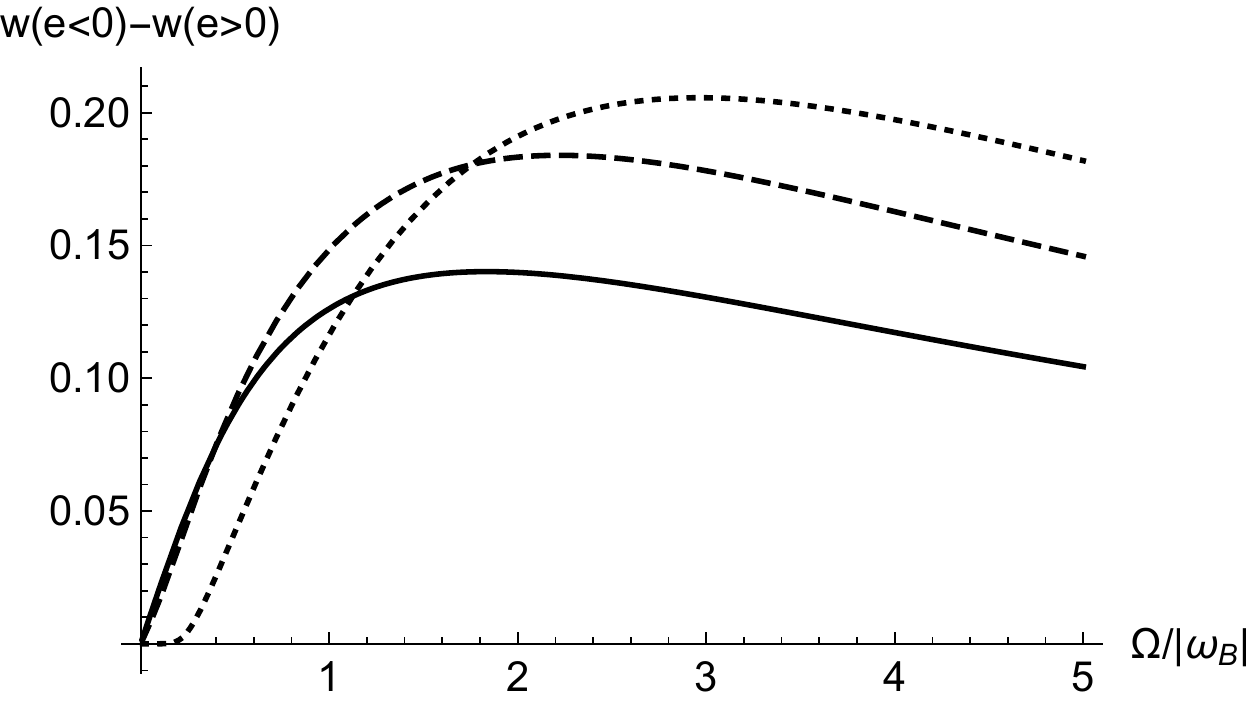} 
             \end{tabular}
  \caption{Left panel: the dissociation probability $w$ as a function of $\Omega/\omega_B$, right panel: the difference between the dissociation probabilities of negative and positive charges. In both panels: $\sqrt{2\e_b/m}=1$, $2m^2/3eB=1$ (hence $\gamma=B/E=1/V$).  Solid line: $V=2/3$, dashed line: $V=1/3$, dotted line: $V=1/6$.}
\label{fig:W}
\end{figure}

For an order of magnitude estimate of the dissociation probability, note that the typical binding of heavy mesons is $2\e_b/m\sim 1$, while $E/B$ is of order of the meson velocity $V$ with respect to plasma. The corresponding dissociation probability is displayed in \fig{fig:W} as a function of $\kappa$ for different $V$'s. One can see that the dissociation probability increases with $V$, which simply reflects the fact that the electric field in the hadron comoving frame is proportional to velocity, as has been previously discussed in \cite{Marasinghe:2011bt}. The effect of rotation is the emergence of the centrifugal force that stimulates the dissociation. \fig{fig:W} indicates that the dissociation probability is a steep function of $\Omega$, hence it should be most pronounced at lower energies, where $\Omega$ is comparable or larger than the synchrotron frequency $|\omega_B|$.

Another conspicuous feature of the dissociation probability is its electric charge dependence. The positive charges, corresponding to $\omega_B>0$ have smaller probability to be torn off the hadron than the negative charges, see the right panel of \fig{fig:W}. In a heavy meson consisting of a heavy and light quarks, it is the electric charge of the light quark that matters. Thus, for example, $D^-$ dissociates more readily than $D^+$. Similarly, $D^0$ containing $\bar u$ quark with electric charge $-2|e|/3$ dissociates with higher probability than  $\bar D^0$ containing $u$ quark with electric charge $2|e|/3$. As a consequence one may observe a smaller number of negatively than  positively charged heavy mesons produced in heavy-ion collisions. This effect is a direct consequence of the collision geometry which requires that the magnetic field and  vorticity point (on average) in the same direction. In a \emph{gedanken} experiment where the direction of the magnetic field and vorticity are opposite, the charge dependence would be opposite.

In this paper it was assumed that the heavy hadron is dragged by the plasma to perform rotational motion with the same vorticity as plasma does, while it moves with finite translation velocity with respect to it. This is of course, not the only reasonable possibility. One can contemplate a situation where the heavy hadron also has a finite rotational velocity with respect to the plasma. In such a case the electromagnetic field in the hadron comoving frame depends on vorticity. This problem can be also handled using the Imaginary Time Method and yields qualitatively similar results. The general conclusion that emerges from examining a few models of electromagnetic fields is that the dissociation probability grows with vorticity and depends on the sign of the electric charge. However, the relative suppression of positive or negative charges  is model-dependent.

\acknowledgments
I  am grateful to Xu-Guang Huang  for sharing with me his results on energy dependence of vorticity in heavy-ion collisions.
This work  was supported in part by the U.S.\ Department of Energy under Grant No.\ DE-FG02-87ER40371.




\begin{thebibliography}{80}

\bibitem{Matsui:1986dk}
T.~Matsui and H.~Satz,
``$\jpsi$ Suppression By Quark - Gluon Plasma Formation,''
Phys.\ Lett.\ B {\bf 178}, 416 (1986).

\bibitem{Mocsy:2013syh}
A.~Mocsy, P.~Petreczky and M.~Strickland,
``Quarkonia in the Quark Gluon Plasma,''
Int. J. Mod. Phys. A \textbf{28}, 1340012 (2013)
[arXiv:1302.2180 [hep-ph]].

\bibitem{Digal:2001ue}
  S.~Digal, P.~Petreczky, H.~Satz,
  ``Quarkonium feed down and sequential suppression,''
  Phys.\ Rev.\  {\bf D64}, 094015 (2001).
  [hep-ph/0106017].

\bibitem{Wong:2004zr}
  C.~Y.~Wong,
  ``Heavy quarkonia in quark gluon plasma,''
  Phys.\ Rev.\  C {\bf 72}, 034906 (2005)
  [arXiv:hep-ph/0408020].


\bibitem{Mocsy:2005qw}
  A.~Mocsy and P.~Petreczky,
  ``Quarkonia correlators above deconfinement,''
  Phys.\ Rev.\  D {\bf 73}, 074007 (2006)
  [arXiv:hep-ph/0512156].

\bibitem{Alberico:2006vw}
  W.~M.~Alberico, A.~Beraudo, A.~De Pace and A.~Molinari,
  ``Quarkonia in the deconfined phase: Effective potentials and lattice
  correlators,''
  Phys.\ Rev.\  D {\bf 75}, 074009 (2007)
  [arXiv:hep-ph/0612062].

\bibitem{Cabrera:2006wh}
  D.~Cabrera and R.~Rapp,
  ``T-matrix approach to quarkonium correlation functions in the QGP,''
  Phys.\ Rev.\  D {\bf 76}, 114506 (2007)
  [arXiv:hep-ph/0611134].

\bibitem{Brambilla:2010cs}
N.~Brambilla, S.~Eidelman, B.~K.~Heltsley, R.~Vogt, G.~T.~Bodwin, E.~Eichten, A.~D.~Frawley, A.~B.~Meyer, R.~E.~Mitchell and V.~Papadimitriou, \textit{et al.}
``Heavy Quarkonium: Progress, Puzzles, and Opportunities,''
Eur. Phys. J. C \textbf{71}, 1534 (2011),
[arXiv:1010.5827 [hep-ph]].


\bibitem{Marasinghe:2011bt}
K.~Marasinghe and K.~Tuchin,
``Quarkonium dissociation in quark-gluon plasma via ionization in magnetic field,''
Phys. Rev. C \textbf{84}, 044908 (2011),
[arXiv:1103.1329 [hep-ph]].


\bibitem{Kharzeev:2007jp}
  D.~E.~Kharzeev, L.~D.~McLerran and H.~J.~Warringa,
  ``The effects of topological charge change in heavy ion collisions: 'Event by
  event P and CP violation',''
  Nucl.\ Phys.\  A {\bf 803}, 227 (2008).
  [arXiv:0711.0950 [hep-ph]].
  
\bibitem{Skokov:2009qp} 
  V.~Skokov, A.~Y.~Illarionov and V.~Toneev,
  ``Estimate of the magnetic field strength in heavy-ion collisions,''
  Int.\ J.\ Mod.\ Phys.\ A {\bf 24}, 5925 (2009),
  [arXiv:0907.1396 [nucl-th]].

\bibitem{Voronyuk:2011jd} 
  V.~Voronyuk, V.~D.~Toneev, W.~Cassing, E.~L.~Bratkovskaya, V.~P.~Konchakovski and S.~A.~Voloshin,
  ``(Electro-)Magnetic field evolution in relativistic heavy-ion collisions,''
  Phys.\ Rev.\ C {\bf 83}, 054911 (2011),
  [arXiv:1103.4239 [nucl-th]].
  
\bibitem{Ou:2011fm} 
  L.~Ou and B.~A.~Li,
  ``Magnetic effects in heavy-ion collisions at intermediate energies,''
  Phys.\ Rev.\ C {\bf 84}, 064605 (2011),
  [arXiv:1107.3192 [nucl-th]].
  
\bibitem{Deng:2012pc} 
  W.~T.~Deng and X.~G.~Huang,
  ``Event-by-event generation of electromagnetic fields in heavy-ion collisions,''
  Phys.\ Rev.\ C {\bf 85}, 044907 (2012),
  [arXiv:1201.5108 [nucl-th]].
  
\bibitem{Tuchin:2013apa} 
  K.~Tuchin,
  ``Time and space dependence of the electromagnetic field in relativistic heavy-ion collisions,''
  Phys.\ Rev.\ C {\bf 88}, no. 2, 024911 (2013),
  [arXiv:1305.5806 [hep-ph]].



\bibitem{Tuchin:2011cg}
K.~Tuchin,
``J/\ensuremath{\psi} dissociation in parity-odd bubbles,''
Phys. Lett. B \textbf{705}, 482-486 (2011),
[arXiv:1105.5360 [nucl-th]].

\bibitem{Machado:2013rta}
C.~S.~Machado, F.~S.~Navarra, E.~G.~de Oliveira, J.~Noronha and M.~Strickland,
``Heavy quarkonium production in a strong magnetic field,''
Phys. Rev. D \textbf{88}, 034009 (2013),
[arXiv:1305.3308 [hep-ph]].

\bibitem{Alford:2013jva}
J.~Alford and M.~Strickland,
``Charmonia and Bottomonia in a Magnetic Field,''
Phys. Rev. D \textbf{88}, 105017 (2013),
[arXiv:1309.3003 [hep-ph]].

\bibitem{Dudal:2014jfa}
D.~Dudal and T.~G.~Mertens,
``Melting of charmonium in a magnetic field from an effective AdS/QCD model,''
Phys. Rev. D \textbf{91}, 086002 (2015),
[arXiv:1410.3297 [hep-th]].

\bibitem{Cho:2014loa}
S.~Cho, K.~Hattori, S.~H.~Lee, K.~Morita and S.~Ozaki,
``Charmonium Spectroscopy in Strong Magnetic Fields by QCD Sum Rules: S-Wave Ground States,''
Phys. Rev. D \textbf{91}, no.4, 045025 (2015),
[arXiv:1411.7675 [hep-ph]].

\bibitem{Sadofyev:2015hxa}
A.~V.~Sadofyev and Y.~Yin,
``The charmonium dissociation in an \textquotedblleft{}anomalous wind\textquotedblright{},''
JHEP \textbf{01}, 052 (2016),
[arXiv:1510.06760 [hep-th]].


\bibitem{Singh:2017nfa}
B.~Singh, L.~Thakur and H.~Mishra,
``Heavy quark complex potential in a strongly magnetized hot QGP medium,''
Phys. Rev. D \textbf{97}, no.9, 096011 (2018),
[arXiv:1711.03071 [hep-ph]].

\bibitem{Dutta:2017pya}
N.~Dutta and S.~Mazumder,
``Majorana flipping of quarkonium spin states in transient magnetic field,''
Eur. Phys. J. C \textbf{78}, no.6, 525 (2018),
[arXiv:1704.04094 [nucl-th]].

\bibitem{Suzuki:2016fof}
K.~Suzuki and S.~H.~Lee,
``Delayed versus accelerated quarkonium formation in a magnetic field,''
Phys. Rev. C \textbf{96}, no.3, 035203 (2017),
[arXiv:1610.09853 [hep-ph]].



\bibitem{Yoshida:2016xgm}
T.~Yoshida and K.~Suzuki,
``Heavy meson spectroscopy under strong magnetic field,''
Phys. Rev. D \textbf{94}, 074043 (2016),
[arXiv:1607.04935 [hep-ph]].

\bibitem{Liu:2018zag}
H.~Liu, X.~Wang, L.~Yu and M.~Huang,
``Neutral and charged scalar mesons, pseudoscalar mesons, and diquarks in magnetic fields,''
Phys. Rev. D \textbf{97}, no.7, 076008 (2018),
[arXiv:1801.02174 [hep-ph]].

\bibitem{Iwasaki:2018pby}
S.~Iwasaki, M.~Oka, K.~Suzuki and T.~Yoshida,
``Hadronic Paschen\textendash{}Back effect,''
Phys. Lett. B \textbf{790}, 71-76 (2019),
[arXiv:1802.04971 [hep-ph]].



\bibitem{Csernai:2013bqa}
L.~P.~Csernai, V.~K.~Magas and D.~J.~Wang,
``Flow Vorticity in Peripheral High Energy Heavy Ion Collisions,''
Phys. Rev. C \textbf{87}, no.3, 034906 (2013)
[arXiv:1302.5310 [nucl-th]].


\bibitem{Csernai:2014ywa}
L.~P.~Csernai, D.~J.~Wang, M.~Bleicher and H.~St\"ocker,
``Vorticity in peripheral collisions at the Facility for Antiproton and Ion Research and at the JINR Nuclotron-based Ion Collider fAcility,''
Phys. Rev. C \textbf{90}, no.2, 021904 (2014)


\bibitem{Becattini:2015ska}
F.~Becattini, G.~Inghirami, V.~Rolando, A.~Beraudo, L.~Del Zanna, A.~De Pace, M.~Nardi, G.~Pagliara and V.~Chandra,
``A study of vorticity formation in high energy nuclear collisions,''
Eur. Phys. J. C \textbf{75}, no.9, 406 (2015)
[erratum: Eur. Phys. J. C \textbf{78}, no.5, 354 (2018)]
[arXiv:1501.04468 [nucl-th]].


\bibitem{Deng:2016gyh}
W.~T.~Deng and X.~G.~Huang,
``Vorticity in Heavy-Ion Collisions,''
Phys. Rev. C \textbf{93}, no.6, 064907 (2016)
[arXiv:1603.06117 [nucl-th]].

\bibitem{Jiang:2016woz}
Y.~Jiang, Z.~W.~Lin and J.~Liao,
``Rotating quark-gluon plasma in relativistic heavy ion collisions,''
Phys. Rev. C \textbf{94}, no.4, 044910 (2016)
[erratum: Phys. Rev. C \textbf{95}, no.4, 049904 (2017)],
[arXiv:1602.06580 [hep-ph]].

\bibitem{Xia:2018tes}
X.~L.~Xia, H.~Li, Z.~B.~Tang and Q.~Wang,
``Probing vorticity structure in heavy-ion collisions by local $\Lambda$ polarization,''
Phys. Rev. C \textbf{98}, 024905 (2018),
[arXiv:1803.00867 [nucl-th]].

\bibitem{Kolomeitsev:2018svb}
E.~E.~Kolomeitsev, V.~D.~Toneev and V.~Voronyuk,
``Vorticity and hyperon polarization at energies available at JINR Nuclotron-based Ion Collider fAcility,''
Phys. Rev. C \textbf{97}, no.6, 064902 (2018),
[arXiv:1801.07610 [nucl-th]].



\bibitem{Deng:2020ygd}
X.~G.~Deng, X.~G.~Huang, Y.~G.~Ma and S.~Zhang,
``Vorticity in low-energy heavy-ion collisions,''
Phys. Rev. C \textbf{101}, no.6, 064908 (2020), 
[arXiv:2001.01371 [nucl-th]].



\bibitem{Chen:2015hfc}
H.~L.~Chen, K.~Fukushima, X.~G.~Huang and K.~Mameda,
``Analogy between rotation and density for Dirac fermions in a magnetic field,''
Phys. Rev. D \textbf{93}, no.10, 104052 (2016)
[arXiv:1512.08974 [hep-ph]].

\bibitem{Mameda:2015ria}
K.~Mameda and A.~Yamamoto,
``Magnetism and rotation in relativistic field theory,''
PTEP \textbf{2016}, no.9, 093B05 (2016),
[arXiv:1504.05826 [hep-th]].


\bibitem{popov-review}
V.~S. Popov, ``Tunnel and multiphoton ionization of atoms and ions in a strong laser field", Physics Uspekhi, {\bf 47}, 855 (2004).

\bibitem{Popov:1997-A}
  V.~S.~Popov, B.~M.~Karnakov and V.~D.~Mur,
 ``Quasiclassical theory of atomic ionization in electric and magnetic fields,''
  Phys.\ Lett.\  A {\bf 229}, 306 (1997).


\bibitem{Popov:1998-A}
  V.~S.~Popov, B.~M.~Karnakov and V.~D.~Mur,
 ``Ionization of atoms in electric and magnetic fields and the imaginary time method",
  JETP {\bf 86}, 860 (1998).
  


\bibitem{Huang:2020xyr}
X.~G.~Huang,
``Vorticity and Spin Polarization --- A Theoretical Perspective,''
[arXiv:2002.07549 [nucl-th]].



  


\end{thebibliography}
\end{document}